\documentclass[a4paper, 10pt, conference]{ieeeconf}      

\IEEEoverridecommandlockouts                              
\makeatletter
\let\NAT@parse\undefined
\makeatother

\usepackage[dvipsnames]{xcolor}

\newcommand*\linkcolours{ForestGreen}

\usepackage{times}
\usepackage{graphicx}
\usepackage{amssymb}
\usepackage{gensymb}
\usepackage{amsmath}
\usepackage{upgreek}
\usepackage{breakurl}

\usepackage{url,hyperref}
\hypersetup{
colorlinks,
linkcolor=\linkcolours,
citecolor=\linkcolours,
filecolor=\linkcolours,
urlcolor=\linkcolours}

\usepackage{algorithm}
\usepackage{algorithmic}

\usepackage[labelfont={bf},font=small]{caption}
\usepackage[none]{hyphenat}

\usepackage{mathtools, cuted}

\usepackage[noadjust, nobreak]{cite}

\usepackage{tabularx}
\usepackage{amsmath}

\usepackage{float}

\usepackage{pifont}

\newcolumntype{Y}{>{\centering\arraybackslash}X}

\usepackage[]{placeins}

\usepackage{placeins}

\usepackage{tikz}

\usepackage[framemethod=tikz]{mdframed}

\usepackage{afterpage}

\usepackage{stfloats}

\title{\LARGE \bf
Monochromatic computed tomography using laboratory-scale setup: proof-of-concept
}

\author{Ari-Pekka Honkanen$^{*1,2}$ 
and Simo Huotari$^{2}$
\\
{\small $^{1}$Comprehensive Cancer Center, Helsinki University Hospital, P.O. BOX 180, FI-00029 HUS,  Finland} \\ {\small $^{2}$Department of Physics, University of Helsinki, P.O. Box 64, FI-00014 Helsinki, Finland}
\thanks{$^*$\texttt{ari-pekka.honkanen@hus.fi}}%
}
\begin{document}

\maketitle
\thispagestyle{empty}
\pagestyle{empty}

\begin{abstract}
In this article, we demonstrate the viability of highly monochromatic full-field X-ray absorption near edge structure based tomography using a laboratory-scale Johann-type X-ray absorption spectrometer based on a conventional X-ray tube source. In this proof-of-concept, by using a phantom embedded with elemental Se, Na$_2$SeO$_3$, and Na$_2$SeO$_4$, we show that the three-dimensional distributions of Se in different oxidation states can be mapped and distinguished from the phantom matrix and each other with absorption edge contrast tomography. 
The presented method allows for volumetric analyses of chemical speciation in mm-scale samples using low-brilliance X-ray sources, and represents a new analytic tool for materials engineering and research in many fields including biology and chemistry.

\end{abstract}

\section{Introduction}

Computed tomography (CT) is a widely used non-destructive method to investigate the three dimensional structure of matter. The clinical CT instruments and a major fraction of laboratory-scale setups are based on polychromatic broad-bandwidth beam produced with conventional X-ray tubes. While this produces a sufficiently high flux of photons for imaging purposes, polychromaticity of the beam has its own drawbacks such as beam-hardening artifacts and insensitivity to the chemical composition of the imaged object. Some amount of chemical contrast can be achieved by dual-energy imaging but the information can be used to separate elements at best into two or three groups based on their atomic number \cite{Rebuffel_2007}. The lack of elemental sensitivity is a significant shortcoming from the viewpoint of materials research as the properties of material rely not only on its elemental composition and distribution but also the chemical speciation of the elements.

These limitations can be overcome with highly monochromatic and tunable X-ray beams such as ones produced with synchrotron and X-ray free electron laser lightsources. One such approach is K-edge subtraction imaging, which has been utilized for example to map the ventilation of airways in lungs during an asthma attack using the xenon gas K-edge absorption imaging \cite{Schlomka_2008, Thomlinson_2018}. 

By adjusting the photon energy of an X-ray beam with $\lesssim$ eV resolution one can even separate the X-ray signals of different chemical species which in turn can be utilized to map the distribution of the species in the sample. 
This method, known as x-ray absorption near-edge spectroscopy (XANES), offers a non-destructive tool for the analysis of the chemistry of a given element, most importantly its oxidation state and local atomic coordination \cite{bunker}. It has shown success in being utilized as a contrast method for full-field tomography in numerous materials research applications such as investigating nano and mesoscale chemical compositions and phase transitions in battery materials \cite{meirer_2006,Kimura_2020, Zhang_2021}, degradation and inactivation of catalyst materials \cite{Gambino_2020, Gao_2021}, and heterogeneity of defect-engineered metal-organic framework crystals \cite{sanchez_2021}. 
It has also been demonstrated that a similar idea can be applied to inelastic X-ray scattering (X-ray Raman spectroscopy) to obtain tomographic data on the chemical state of low-Z elements to e.g. spatially distinguish $sp^2$ and $sp^3$ bonds in carbon materials \cite{Huotari_2011}.

The aforementioned techniques require a highly brilliant, energy-tunable X-ray light source, such as a synchrotron light source, which limits their applicability in the laboratory scale. However, due to high demand and scarcity of beamtime at large scale synchrotron and X-ray free electron laser lightsources, the laboratory-scale X-ray spectrometry has experienced a renaissance in the recent years. Despite their orders of magnitude lower photon output, laboratory-scale instruments have proven to be a viable alternative to large-scale facilities in many applications \cite{Mortensen_2017, Wang_2017, Lusa_2019, Moya_Cancino_2019,Mottram_2020, Zimmermann_2020}.

\begin{figure}[ht]
\centering
\includegraphics[width=0.9\columnwidth]{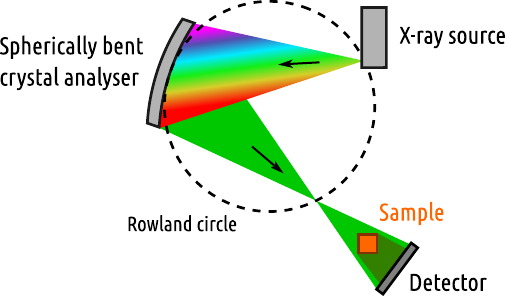}
\caption{The schematic drawing of the XAS-CT setup. Polychromatic X-rays produced by the X-ray tube are monochromatised with the spherically bent crystal analyser. The sample to be imaged is illuminated by the monochromatised beam by moving it away from the Rowland circle so that the defocused beam covers it completely. The beam transmitted through the sample is recorded with a position-sensitive detector. \label{fig:schema}}
\end{figure}

In our previous work \cite{Honkanen_2019}, we demonstrated chemically sensitive 2D-imaging using a Johann-type X-ray absorption (2D-XANES) spectrometer based on a conventional X-ray tube as presented in Fig.~\ref{fig:schema}. The polychromatic beam of the primary source is directed at a spherically bent crystal analyser which monochromatises and refocuses the beam at on the Rowland circle which acts as a secondary source. The sample and the imaging detector are set downstream from the secondary focus. Chemical sensitivity is obtained by adjusting the energy of the diffracted photons and recording the spatially resolved changes in the attenuation coefficients.

\begin{figure}[ht]
\centering
\includegraphics[width=\columnwidth]{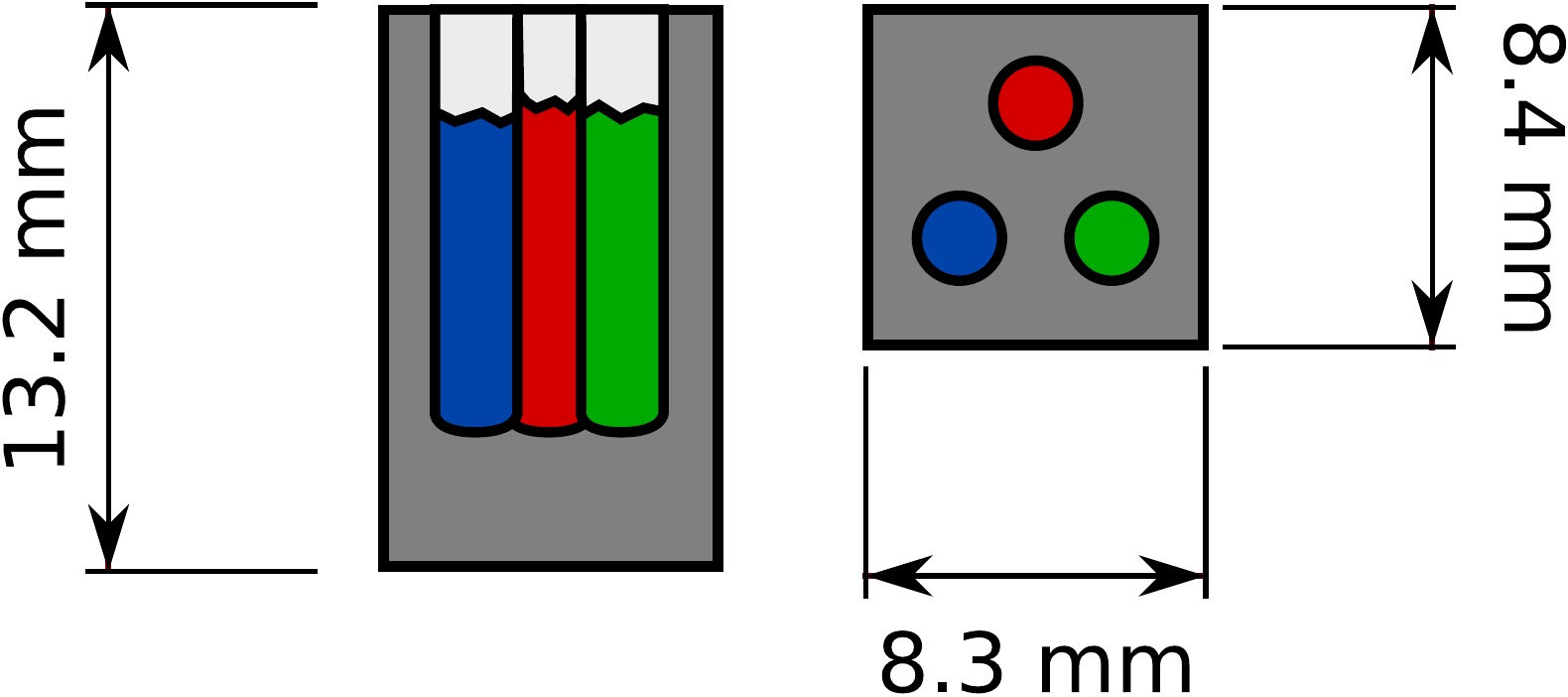}
\caption{Schematic drawing of the phantom. The cuboid PMMA (dark grey) was drilled with three holes each of which was filled with a mixture of a selenium compound (elemental Se (red), Na$_2$SeO$_3$ (green) and Na$_2$SeO$_4$(blue)) and starch. The holes were capped with tissue paper (light grey).  \label{fig:phantom}}
\end{figure}

In this work we develop the imaging capabilities of such a laboratory setup further by demonstrating 3D imaging with chemical contrast (3D-XANES) using the low-brilliance x-ray source. We prepared a PMMA phantom (Fig.~\ref{fig:phantom}) which was embedded with Se in different chemical states and show that mapping the 3D spatial distribution of different chemical species is viable using the setup described.

\section{Results}

Fig.~\ref{fig:schema} shows the schematic design of the experiment. The polychromatic beam is produced by a conventional 1.5~kW X-ray tube with an Ag anode. The beam is monochromatised and refocused on the focal spot using a spherically bent strip-bent Si(953) crystal analyser with a bending radius of 0.5~m \cite{Rovezzi_2017}. The focal point of the crystal analyzer produces a secondary source for the cone-beam imaging, with sample located downstream, and the image of which is captured by a position-sensitive photon-counting MiniPIX detector by Advacam, which is based on a TimePIX \cite{Llopart_2007} direct conversion Si detector with $256\times256$ square pixels with the side length of 55~$\upmu$m.
To obtain a spectrum, the crystal analyzer Bragg angle is scanned across the relevant range (in this case, 77.47$^\circ$ to 73.01$^\circ$, corresponding to a range of 12.54 to 12.80 keV). The sample and the detector follow the moving monochromatic cone with high precision using motorized stages. 

\begin{figure}[ht]
\centering
\includegraphics[width=\columnwidth]{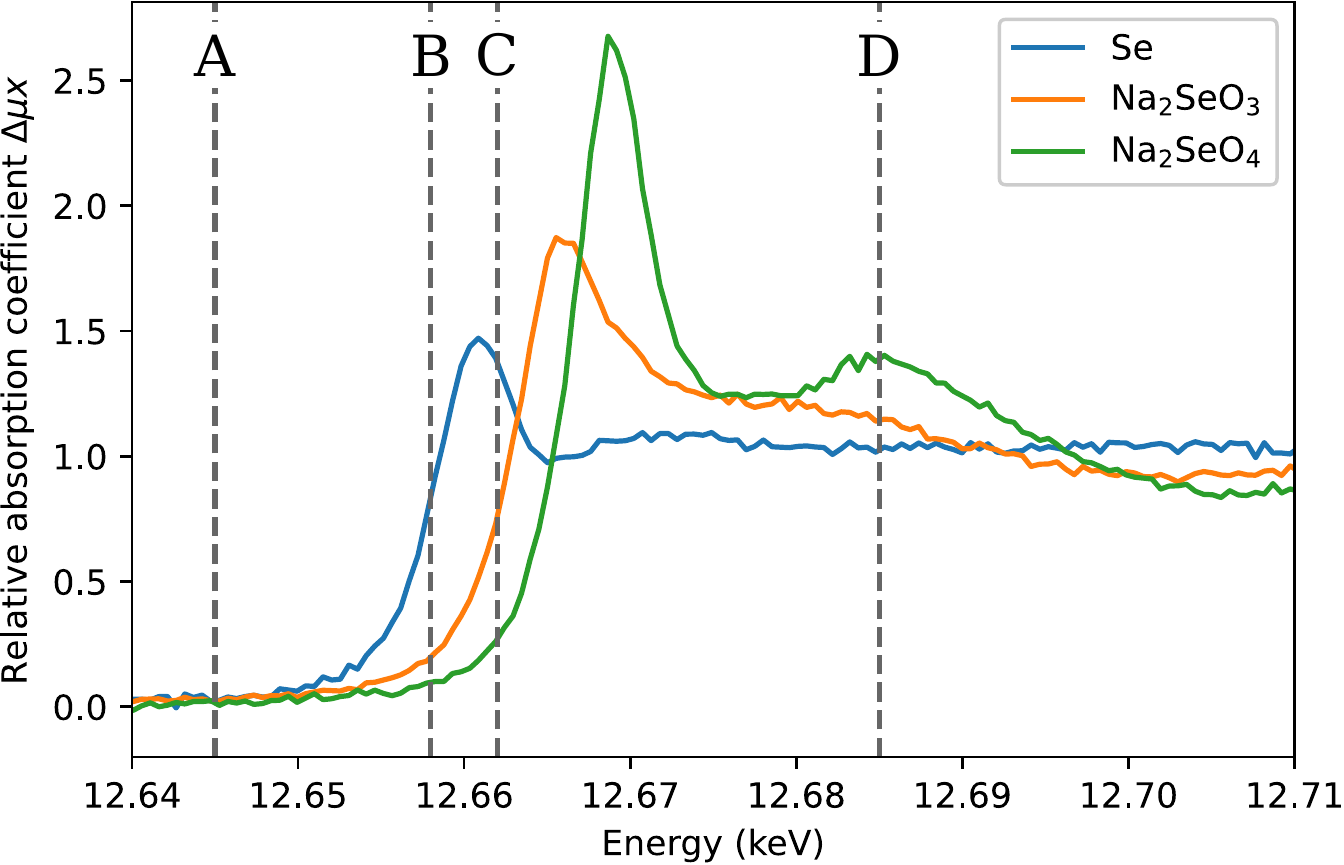}
\caption{Background-subtracted and normalized K-edge absorption spectra of elemental Se, Na$_2$SeO$_3$ [Se(IV)] and Na$_2$SeO$_4$ [Se(VI)]. The vertical lines indicate the acquisition photon energies of 12.645 (A), 12.658 (B), 12.662 (C), and 12.685 (D) keV. \label{fig:Se_spectra}}
\end{figure}

The contrast of Se-K edge XANES is illustrated in Fig. \ref{fig:Se_spectra}. It shows the relative increase of the attenuation coefficient at the photon energy range that corresponds to the Se K near edge region. The background of the attenuation owing to other electrons than the Se $1s$ have been subtracted for photon energies below the K edge, and the background-subtracted spectra have been normalized to an equal area in the energy region of the plot. The CT scans were obtained at the photon energies labelled A-D. As will be shown below, sampling the spectra at these four distinct energies is sufficient to distinguish the three different oxidation states of Se that are used in this experiment.

For each monochromated photon energy, a transmission (projection) image is acquired for various sample orientations at 1.8$^{\circ}$ intervals, with a 100 s exposure time.

\begin{figure}[ht]
\centering
\includegraphics[width=\columnwidth]{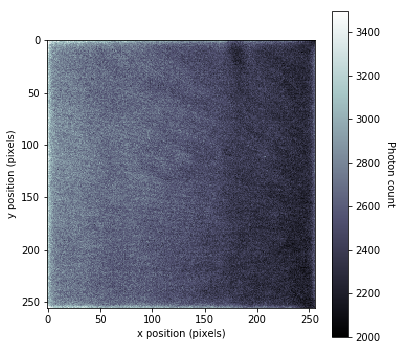}
\caption{Flat field image with a monochromatised beam. Defunct pixels are removed from the image by replacing them with the median value of the neighbouring pixels. \label{fig:flat_field}}
\end{figure}

The first step in the data analysis is to calibrate the instrument via, e.g., the flat field image.
An example of a single flat field image is presented in Fig.~\ref{fig:flat_field}.
The image is not completely uniform but it exhibits a horizontal slope in the intensity. Since the electron beam inside the X-ray tube propagates horizontally as well, the observed slope could be due to the intensity variation in the source beam due to the anode heel effect. Non-uniform stray scattering inside the chamber may also contribute to the observed gradient.
With better collimation, shielding of the equipment and use of vacuum or He-filled chamber to reduce the scatter, this non-constant background probably can be partly reduced in the future. Slightly darker vertical lines at $x=60$ and $x=180$ px are due to the gaps between the strips of the crystal wafer analyser \cite{Rovezzi_2017}. The mean count rate of photons per pixel was 2.6~ph~s$^-1$~px$^-1$. Almost all photon count values fall between the range 2.2--3.0~ph~s$^-1$~px$^-1$ or approximately within $\pm 15~\%$ of the mean value.

\begin{figure}[ht]
\centering
\includegraphics[width=\columnwidth]{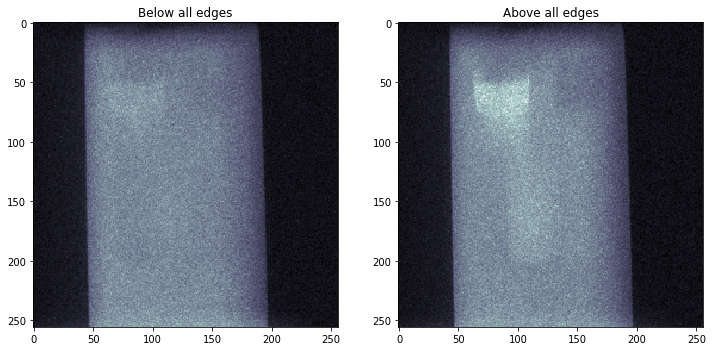}
\caption{Projections of the phantom taken at energies A (below all edges, left) and D (above all edges, right) with the flat field correction.\label{fig:projections}}
\end{figure}

In Fig.~\ref{fig:projections} two projections are presented, taken at energy A and D, respectively.
As expected, there is virtually no difference in the integrated attenuation coefficient of the phantom except at the locations which contain Se. The resolution of the image was calculated from the projection at energy A by differentiating the edge of the phantom row-by-row at the lower left corner of the projection, fitting Gaussian functions to the resulting line spread functions, and taking the mean of the FWHM (full width at half maximum) and its standard error of the fitted curves. The spatial resolution of the projection was found to be $85\pm4$~$\upmu$m, or $1.6\pm0.1$~px.

\begin{figure}[ht]
\centering
\includegraphics[width=\columnwidth]{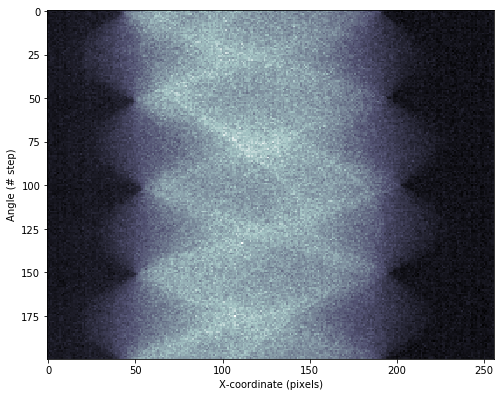}
\caption{Example of a sinogram from the projections taken at energy A \label{fig:sinogram}}
\end{figure}

An example sinogram is presented in Fig.~\ref{fig:sinogram}. As can be seen in the sinogram and projections in Fig.~\ref{fig:projections}, the background level of the signal is slightly higher on the left side of the images compared to the right hand side despite the flat field correction. The reason for this is unclear but it could be possibly due to diffuse scattering and fluorescence from the irradiated parts inside the instrument enclosure. The uneven background was corrected by perfoming a linear fit to it and subtracting the fit from the sinograms before reconstruction.

\begin{figure}[ht]
\centering
\includegraphics[width=\columnwidth]{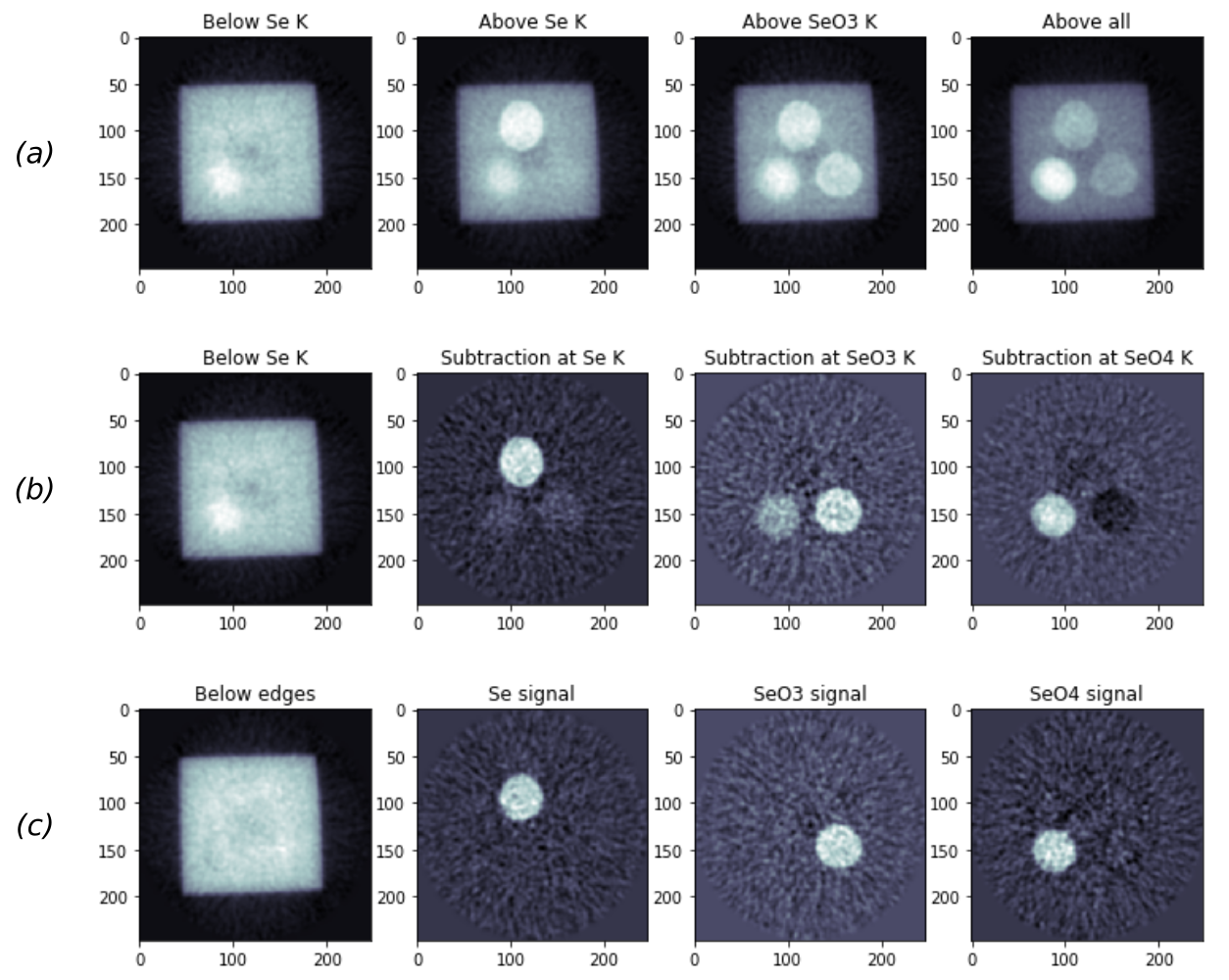}
\caption{\emph{(a)} Axial slices of the reconstructed tomograms acquired at energies A--D respectively from left to right; \emph{(b)} Separated Se signals by subtracting the tomograms with adjacent energies. From left to right, the first image is equal to the original tomogram at energy A, and the following ones are the differences B$-$A, C$-$B, and D$-$C, respectively; \emph{(c)} Separated signals using matrix inversion for the phantom without Se, Se(0), Se(IV), and Se(VI).
\label{fig:tomograms}}
\end{figure}

The tomograms were reconstructed using the filtered back projection algorithm assuming the parallel beam geometry. Slices of tomograms acquired at each photon energy are presented in Fig.~\ref{fig:tomograms} a), where strong energy-dependent variation can be seen in the attenuation in the parts containing selenium. 
Similar to the projection images, the resolution of the reconstruction was determined from the top edge of the reconstructed slice at energy A using Gaussian fitting and was found to be $226 \pm 5$~$\upmu$m, or $4.1\pm0.4$~px.

In the usual case of K-edge imaging, where the fine structure of the edge can be ignored, the signal containing the abrupt change in the attenuation coefficient can be separated from the rest via simple subtraction. However, as seen in Fig.~\ref{fig:tomograms} b), whereas the rest of the phantom can be cleanly removed from the Se signals by subtracting the adjacent lower energy point from the higher one, the signals from different Se species can not be fully separated from each other by such a simple subtraction procedure.
In order to perform a proper separation, we note that since a CT image is a direct measure of the linear attenuation coefficient, a set of reconstructed tomograms $\mathbf{y}$ taken at different photon energies can be written as
$\mathbf{y} = \mathcal{C}\mathbf{x}$, where $\mathbf{x}$ contains the different components of the imaged object and $\mathcal{C}$ contains the mass attenuation coefficients of the components at measured energies. When $\mathcal{C}$ is known, $\mathbf{x}$ can be solved from $\mathbf{x} =  \mathcal{C}^{-1} \mathbf{y}$, where $\mathcal{C}^{-1}$ is the (pseudo)inverse of $\mathcal{C}$. Since in this phantom the different species of Se are spatially well separated, $\mathcal{C}$ can be determined (up to multiplicative factors) directly from the tomograms by summing the pixel values over arbitrarily chosen ROIs covering the different Se samples; in a more convoluted case the attenuation coefficients could be measured from known reference samples. When $\mathcal{C}$ can be determined accurately, the matrix inversion can fully separate different signal components as shown in Figure~\ref{fig:tomograms} c).

\begin{figure}[ht]
\centering
\includegraphics[width=\columnwidth]{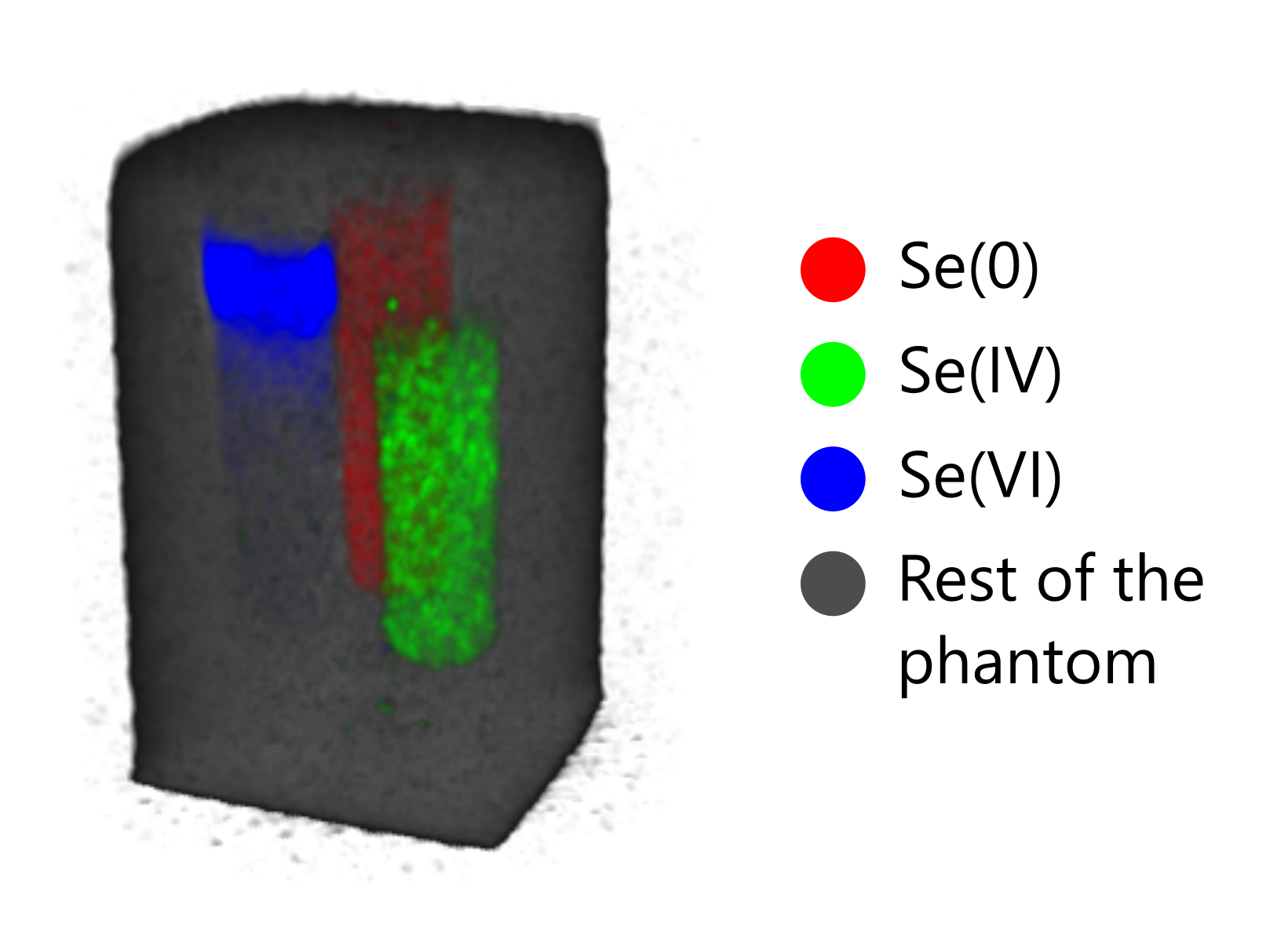}
\caption{3D visualisation of different chemical species of Se inside the phantom. The rendering was made with 3D Slicer 4 \cite{slicer_website, Fedorov_2012}. \label{fig:3d_reco}}
\end{figure}

Figure~\ref{fig:3d_reco} shows the 3D distribution of different species of Se inside the phantom. Whereas Se(0) and Si(IV) are found to be evenly distributed in boron nitride, Se(VI) is found to be heavily concentrated towards the top part of the container. The most likely explanation to the inhomogeneity comes from the preparation of the phantom. Na$_2$SO$_4$ was mixed in boron nitride using ethanol to aid the blending. Apparently ethanol had not completely evaporated before the mixture was placed in the phantom which caused it and the dissolved  Se(VI) to be absorbed into the tissue paper that was used to seal the container. Albeit unexpected, chemical contrast CT imaging  has thus revealed new information about the composition of the phantom, a kind of which would not have been possible to obtain by visual inspection.

\section{Discussion}

The proof-of-concept study presented above demonstrates the feasibility of chemical contrast CT imaging using a laboratory-scale setup. Due to the relatively low flux of photons, the measurement times are long but not impractically so for the purposes of materials research where the concentrations of elements of interest are often high in terms of their X-ray absorption strength.

In this work we used \emph{a priori} knowledge of the distribution of the Se species in the phantom to decompose the absorption signals to chemical mapping. However, since the XAS instrument is naturally equipped with spectral recording capabilities, one could measure the needed attenuation coefficients as a function of energy from known reference samples which could be used to perform the matrix inversion of an unknown sample. Alternatively other decomposition methods such as principal component analysis and non-negative matrix factorisation could be used.

The reconstructions in this work were performed with the filtered backscattering method but due to the low photon count, the image quality of the tomograms of the presented setup could significantly benefit from more advanced reconstruction algorithms. Algorithms such as total variation regularization could also be used to obtain full spatially resolved spectra because, analogous to 4D scans, tomograms taken at different energies have relatively little variation between them which suggests that higher quality reconstructions could be achieved when the information in the whole dataset is taken into consideration simultaneously.

Finally, as it is true with more typical CT setups, the geometric magnification could be utilized with XAS-CT as well to overcome the pixel size limitations of the detector. This could be achieved even without a micro-focus tube because the secondary source size could be adjusted with appropriate slitting. However, the technical feasibility of this requires further work because slitting lowers the photon output further and slitting needs to accommodate the movement of the secondary focus. In addition, the spherical astigmatism separates the vertical and horizontal secondary foci from each other which could cause additional issues; however, this could be solved with toroidally bent crystal analysers.

\section{Conclusions}

We have described and demonstrated the viability of highly monochromatic laboratory-scale CT imaging using an X-ray absorption spectrometer based on a conventional X-ray tube lightsource and spherically bent crystal analysers. By accurately tuning the energy of the monochromatised beam, we were able to obtain a distinct contrast difference between different chemical species of Se inside a PMMA phantom. Whereas more research is needed to improve and further characterise practical implementations of laboratory-based XAS-CT, we anticipate that incorporation of monochromatic imaging and CT to the laboratory-XAS instruments will provide a valuable addition to the growing toolbox of laboratory-based X-ray analysis methods. Higher spatial and energy resolutions can be achieved using synchrotron-light sources, but the fact that laboratory-scale experiments can often be performed at will and as long-term campaigns with abundant access to the instrumentation, highlights the extremely useful nature of the laboratory-scale approach.


\section{Methods}

\subsection{Experimental setup}
The CT setup is based on the Johann-type laboratory-scale XAS instrument previously described in \cite{Honkanen_2019} and its schematic drawing is presented in Fig.~\ref{fig:schema}.  The electron beam spot size is $0.4 \times 8$~mm which at 6$^{\circ}$ take-off angle projects to the primary X-ray source size of $0.4 \times 0.8$~mm (V $\times$ H). The analyser focuses the monochromatized beam to a focal "point" at the Rowland circle (spherical astigmatism separates the vertical and horizontal foci)  which acts as a secondary source. The sample to be imaged is moved downstream away from the secondary focus so that the divergent beam illuminates it completely. The sample was rotated using a stepper motor fitted on a custom-built stage set on the linear translation stage.

\subsection{Experiment}
\begin{figure}[ht]
\centering
\includegraphics[width=\columnwidth]{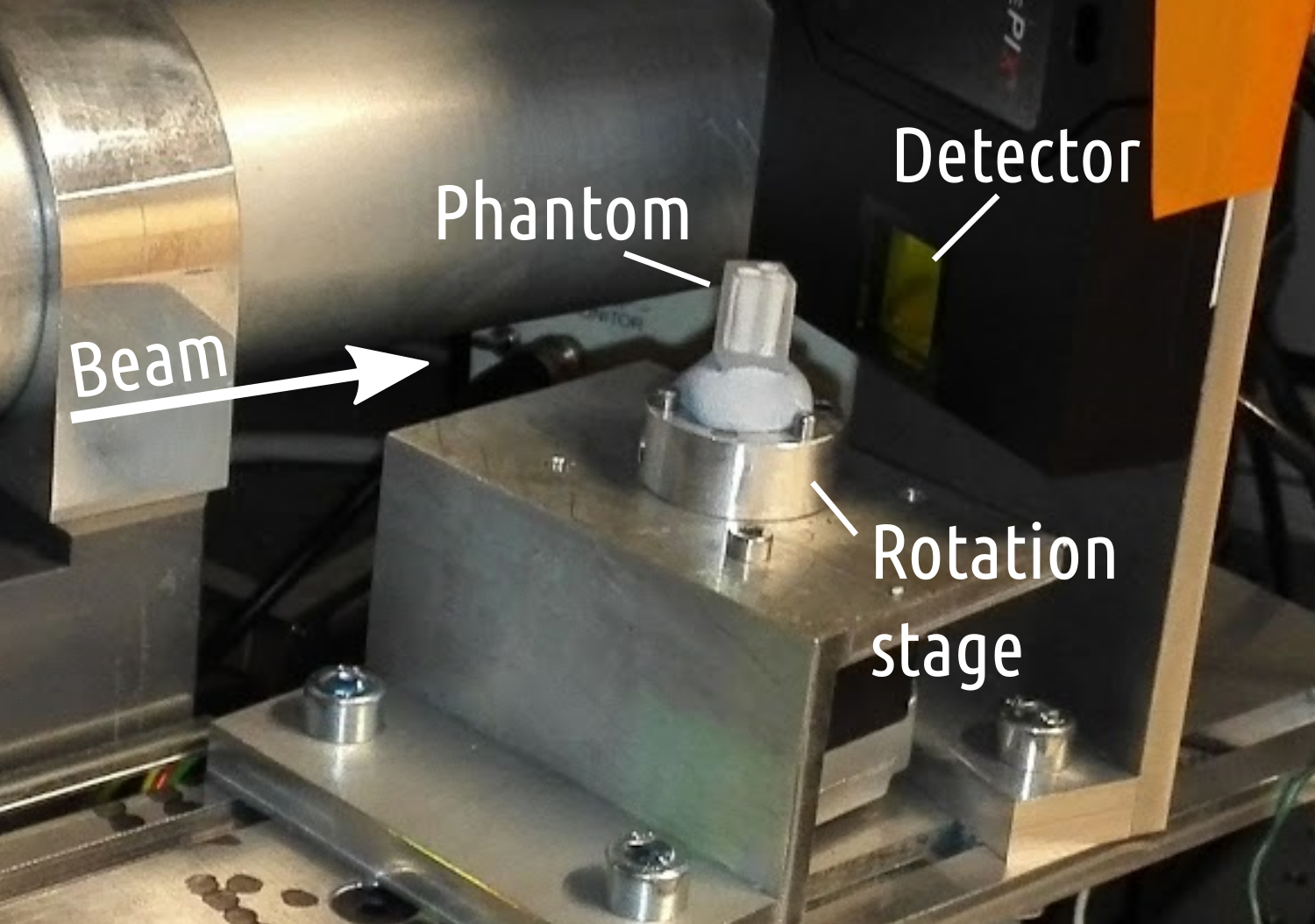}
\caption{Photograph of the tomography setup. The detector in the picture is a different TimePIX-based model than the one used in this work.\label{fig:setup}}
\end{figure}

We tested the technique by imaging a selenium containing phantom (Figs.~\ref{fig:phantom} and \ref{fig:setup}). The phantom consisted of a cuboid piece of PMMA with the side lengths of $8.2 \times 8.3 \times 13.2$~mm. Parallel to the long axis of the phantom, three holes of approximately $2$~mm in diameter and $9.6$~mm in depth were drilled. Each hole were filled respectively with a mixture of boron nitride and one of the three different chemical species of selenium: elemental Se (black allotrope), Na$_2$SeO$_3$, and Na$_2$SeO$_4$. For the rest of the work, these chemical species will be referred to by their nominal oxidation states, \emph{viz.} Se(0), Se(IV), and Se(VI), respectively. The amount of each Se compound was measured so that the absorption coefficient above the Se K absorption edge was $\mu x \approx 1$. With 2~mm of unfilled margin left for each hole, in terms of sample mass this corresponds to approx. 0.7~mg of elemental Se. The filled holes were capped with tissue paper.

The projections of the phantom were acquired at four different photon energies near Se K-edge: 12.645 (A), 12.658 (B), 12.662 (C), and 12.685 (D) keV. The energies were chosen so that A was well below all the absorption edges of Se compounds, B was above the white line of Se(0) but below that of Se(IV), C was above the white line of Se(IV) but below that of Se(VI), and D was well above all the absorption edges. The relationship between the acquisition energies and the absorption spectra are presented in Figure~\ref{fig:Se_spectra} 

At each photon energy 200 projections were acquired over 360$^{\circ}$ rotation of the sample at 1.8$^{\circ}$ intervals. Each projection were exposed for 100 s. Before and after the acquisition of the projections 2 flat field images were recorded without the sample for 1000~s each. Due to extremely low background noise of the detector, no dark field images were taken. Si(953) analyser was used as a monochromator and the acceleration voltage and current of the X-ray tube were 20~kV and 40~mA, respectively. With this combination of acceleration voltage and crystal reflection only the lowest order diffraction peak is present. The mean primary source to analyser distance was approx. 48.4~cm (Bragg angles 74.8$^{\circ}$--75.5$^{\circ}$) and secondary source to detector distance was approx. 21~cm.

\bibliographystyle{ieeetr}
\bibliography{main}

\section*{Author contributions}

\noindent APH designed and performed the experiments and data-analysis. The article was written by APH and SH. The figures were prepared by APH.

\section*{Acknowledgements}

\noindent The authors want to thank Dr. Heikki Suhonen for the feedback on the CT reconstruction. APH was funded by the University of Helsinki Doctoral Program in Materials Research and Nanosciences (MATRENA). SH was supported by the Academy of Finland grant 295696. We thank the Helsinki Center for X-ray Spectroscopy for provision of beamtime with the Hel-XAS spectrometer.

\clearpage

\end{document}